\documentclass[
 prl,
 reprint,
 floatfix,
 amsmath,amssymb,
 showpacs,
 superscriptaddress,
 citeautoscript,
 nonatbib
]{revtex4-1}
\usepackage{xr}
\makeatletter
\newcommand*{\addFileDependency}[1]{
  \typeout{(#1)}
  \@addtofilelist{#1}
  \IfFileExists{#1}{}{\typeout{No file #1.}}
}
\makeatother

\usepackage{amssymb}

\usepackage[breaklinks]{hyperref}%hyperef package does internal and external linking (labels, urls)
\usepackage[all]{hypcap}%addition to hyperref, it ensures figure is correctly displayed (and not only the caption) when link is clicked
\hypersetup{
    plainpages=false,
    unicode=false,                          % non-Latin characters in AcrobatĂŻÂżÂ˝s bookmarks
    pdfborder={0 0 0},
    pdftoolbar=true,                        % show AcrobatĂŻÂżÂ˝s toolbar?
    pdfmenubar=true,                        % show AcrobatĂŻÂżÂ˝s menu?https://www.overleaf.com/13403665mnpkhvmzmbnjhttps://www.overleaf.com/13403665mnpkhvmzmbnj
    pdffitwindow=false,                     % window fit to page when opened
    pdfstartview={FitH},                    % fits the width of the page to the window
    pdftitle={},                            % title
    pdfauthor={},                           % author
    pdfproducer={LNCMI-CNRS-UGA-UPS-INSA},  % producer of the document
    pdfkeywords={High} {Magnetic} {Fields}, % list of keywords
    pdfnewwindow=true,                      % links in new window
    linktoc=section,
    colorlinks=true,                        % false: boxed links; true: colored links
    linkcolor=blue,                         % color of internal links
    citecolor=red,                          % color of links to bibliography
    filecolor=magenta,                      % color of file links
    urlcolor=blue                           % color of external links
}

\usepackage{graphicx}% Needed to include png/jpg/pdf figure files
\usepackage{float}
\usepackage{amsmath}
\usepackage{bm}
\usepackage{color, soul}
\usepackage[normalem]{ulem}
\usepackage{siunitx}
\usepackage{upgreek}
\usepackage[usenames,dvipsnames]{xcolor}
\usepackage{textcomp}
\usepackage{chemformula}
\usepackage[normalem]{ulem}
\usepackage{rotating}
\usepackage{booktabs}
\usepackage{amssymb}
\usepackage{colortbl}   %to fill rows color

\definecolor{darkgreen}{HTML}{007F00}
\colorlet{cyan}{darkgreen}

\begin{document}

\author{Xiangzhou Zhu}
\affiliation{Physics Department, TUM School of Natural Sciences, Technical University of Munich, 85748 Garching, Germany}

\author{David A. Egger}
\email{david.egger@tum.de}
  \affiliation{Physics Department, TUM School of Natural Sciences, Technical University of Munich, 85748 Garching, Germany}

\title{The Effect of Overdamped Phonons on the Fundamental Band Gap of Perovskites}

\date{\today}% It is always \today, today,

\begin{abstract}
Anharmonic atomic motions can strongly influence the optoelectronic properties of materials but how these effects are connected to the underlying phonon band structure is not understood well. 
We investigate how the electronic band gap is influenced by overdamped phonons, which occur in an intriguing regime of phonon-phonon interactions where vibrational lifetimes fall below one oscillation period. 
We contrast the anharmonic halide perovskite CsPbBr$_3$, known to exhibit overdamped phonons in its cubic phase, with the anharmonic oxide perovskite SrTiO$_3$ where the phonons are underdamped at sufficiently high temperatures.
Our results show that overdamped phonons strongly impact the band gap and cause slow dynamic fluctuations of electronic levels that have been implicated in the unique optoelectronic properties of halide perovskites. 
This finding is enabled by developing augmented stochastic Monte Carlo methods accounting for phonon renormalization and imaginary modes that are typically neglected. 
{Our work provides guidelines for capturing anharmonic effects in theoretical calculations of materials.}
\end{abstract}

\maketitle
Atomic vibrations play an important role for condensed-phase systems and their properties at finite temperatures \cite{dove1993}.
Relevant for applications in optoelectronic devices, the electronic structure of crystalline materials can be strongly influenced by these vibrations. 
For example, they are known to dictate the temperature-dependence of carrier transport in classical inorganic semiconductors such as Si and GaAs \cite{yu2010}.

The starting point to include effects of atomic vibrations on optoelectronic properties is the harmonic approximation and the phonon picture.
It results in coherent atomic oscillations, commonly dubbed lattice vibrations, which perturb the electronic levels compared to the  perfectly ordered and static crystal.
From a theoretical perspective, phonon-induced changes of the electronic structure can be captured in the framework of electron-phonon interactions that was established with many-body perturbation theory \cite{mahan2000}.
Combinations of this framework with first-principles calculations were extensively developed and enable accurate finite-temperature calculations of materials \cite{Giustino2017a}.
A canonical method to calculate thermal renormalization of electronic levels is Allen–Heine–Cardona theory \cite{allen1976,allen1981}.
It rests on several assumptions including the adiabatic approximation, the aforementioned harmonic approximation and treating electron-phonon interactions as a small perturbation. 
While recent studies showed that this well-established theory is not accurate enough to describe the band gap of perovskites at finite temperature \cite{saidi2016a,wu2020b}, it is unclear precisely which of the involved approximations is responsible for it.

Relevant in this context are anharmonic lattice vibrations, which arise from the presence of higher-order terms in the potential energy surface beyond the quadratic one of the harmonic approximation.
A significance of anharmonicity implies that atomic displacements at finite temperature are large enough such that atoms visit these higher-order regions.
Recent studies showed that anharmonic lattice vibrations can strongly influence the optoelectronic properties of diverse crystalline materials. 
Examples include the optical gap of solid hydrogen \cite{monacelli2021a}, band gaps of organic \cite{alvertis2022} and nitride semiconductors \cite{hegner2024}, excitonic properties in chalcogenide quantum wells \cite{kastl2023} and various properties of halide 
\cite{quarti2016,marronnier2017,yaffe2017,marronnier2018,klarbring2020,gehrmann2022b,seidl2023,schilcher2023a,zacharias2023b} and oxide perovskites \cite{zhou2018,zhou2019a,zacharias2020b}.
Hence, phonon-phonon interactions impact electron-phonon interactions and optoelectronic properties of various materials. 

However, anharmonicity can manifest in several different ways \cite{cohen2022,dove1993} and it is a priori unclear how different manifestations of anharmonicity impact electron-phonon interactions and optoelectronic properties. 
This lack of conceptual understanding hampers finite-temperature calculations of electronic properties of materials from first-principles.
For example, thermal volume expansion due to anharmonicity can be straightforwardly integrated in first-principles calculations by calculating the electron-phonon interactions in the quasi harmonic approximation \cite{dove1993}.
More profound anharmonic effects occur when instantaneous atomic structures strongly deviate from the idealized crystal structure. 
These effects required extending perturbative treatments of the electron-phonon interactions, e.g., by finite-temperature renormalization of phonons \cite{wu2020b,patrick2015e} or specific quasi-random structures and displacements \cite{zacharias2015,zhao2020,zacharias2023a}.
However, anharmonic atomic motions can lead to a breakdown of the phonon quasiparticle picture altogether when phonon-phonon interactions are very strong. 
This scenario raises puzzling questions of how optoelectronic properties of materials are influenced by such dramatic anharmonic effects and how to include them using first-principles calculations. 
As a case in point, a recent study applying constrained density functional theory (DFT) with the stochastic self-consistent harmonic approximation found an intriguing coupling between anharmonic and optoelectronic effects triggering photoinduced phase transitions in GeTe \cite{furci2024}.

In this Letter, we aim to disentangle the impact of various phonon-induced effects on the fundamental band gap of perovskites. 
Measuring the energy between the valence band maximum and conduction band minimum, the band gap is perhaps the most important optoelectronic property of semiconductors and insulators. 
Starting from a first-principles Monte Carlo (MC) method \cite{zacharias2015} rooted in the harmonic approximation, we augment it by including the effects of phonon renormalization and imaginary modes that are typically neglected in standard treatments. 
Our multi-level first-principles approach contrasts the MC results with those obtained using molecular dynamics (MD). 
The latter serves as a benchmark because DFT-MD calculations capture anharmonic effects incorporating all orders of phonon-phonon interactions in a non-perturbative way. 
This allows us to address the importance of overdamped phonons for the band gap, i.e., lattice vibrations with a lifetime dropping below their oscillation period, in two exemplary perovskites. 
We find that slower atomic dynamics are due to overdamped phonons and determine the large thermal renormalization of the band gap. 
{We also show that these slower oscillations are due to strong phonon-phonon interactions and neglected in current perturbative treatments.} 
This work delivers microscopic insight on the role of overdamped phonons highlighted in several recent studies  \cite{ferreira2020a,stock2020,weadock2020,lanigan-atkins2021b,weadock2023b,fransson2023} for optoelectronic properties and provides guidelines of how to include these effects in first-principles calculations. 

We start our investigation of the impact of phonons on the band gap for cubic SrTiO$_3$ (STO) at \textit{T}=1000\,K.
STO exhibits anharmonic vibrations \cite{tadano2015,he2020} that were reported to impact its optoelectronic properties \cite{zhou2018,zhou2019a,wu2020b,zacharias2020b}.
However, its phonons are known to be underdamped at the temperature considered here \cite{shapiro1972,bruce1983,shirane1969},
i.e., their lifetimes are large compared to oscillation periods and the phonon quasiparticle picture remains valid.
We combine various methods of computing the finite-temperature band gap with first-principles DFT calculations {and non-perturbative treatments of the electron-phonon interactions, }using \texttt{VASP} \cite{kresse1996} and the PBEsol functional \cite{perdew2008}, see Table~\ref{table:1} \cite{sm2024}.
Starting from the band gap of the static structure (1.81\,eV), we perform stochastic MC calculations displacing atoms according to harmonic phonons calculated for cubic STO (MC$_\mathrm{harm}$ in Table~\ref{table:1}). 
This procedure finds a band-gap opening of roughly 0.2\,eV and is in stark contrast with the experimentally established reduction of the band gap with increasing temperature \cite{kok2015}.

\begin{table}
\caption{Comparison of band gaps of SrTiO$_3$ at 1000\,K and CsPbBr$_3$ at 425\,K calculated using density functional theory (DFT) and a static structure (static), a Monte Carlo method with harmonic (MC$_\mathrm{harm}$) and renormalized phonons (MC$_\mathrm{renorm}$), and molecular dynamics (MD).}
\begin{ruledtabular}
    \begin{tabular}{l c c c c}
    & \multicolumn{4}{c}{Band gap (eV)} \\\hline
     & {Static} & {MC$_\mathrm{harm}$} & {MC$_\mathrm{renorm}$ } & {MD} \\
    \hline
    SrTiO$_3$ & 1.81 & 1.97 & 1.47 & 1.54 \\
    CsPbBr$_3$ & 1.43 & 1.78 & 1.64 & 2.15 \\
    \end{tabular}
\end{ruledtabular}
\label{table:1}
\end{table}

Recent theoretical work showed that the correct result can be obtained once renormalized phonons are included when calculating the band gap of STO at finite temperature \cite{wu2020b}.
This is primarily because the electron-phonon interactions change because of phonon renormalization\cite{wu2020b}.
We confirm this finding by computing the renormalized phonon dispersion of STO at \textit{T}=1000\,K using MD and \texttt{TDEP} \cite{hellman2013a,knoop2024,sm2024}.
Including the effect of phonon renormalization in our MC treatment (MC$_\mathrm{renorm}$ in Table~\ref{table:1}), we find a band gap value of 1.47\,eV that is reduces compared to the value of the static structure as expected. 
Finally, we use the same DFT approach but calculate the band gap as a the mean value from MD snapshots in order to include all phonon-phonon {interactions in a non-perturbative way}. 
This procedure finds a similar value as the MC method including renormalized phonons (see Table~\ref{table:1}). 
For anharmonic materials exhibiting underdamped phonons, {perturbatively treating thermal renormalization of phonons and including it in electron-phonon interactions} leads to accurate thermal trends of the band gap.

We now consider the halide perovskite CsPbBr$_3$ (CPB) in its cubic phase at \textit{T}=425\,K.
Like STO, CPB is known to exhibit anharmonic effects at this temperature \cite{yaffe2017,lanigan-atkins2021b,gehrmann2022b,tadano2022a,seidl2023,fransson2023,zacharias2023b,baldwin2024}.
Contrary to STO, it is well-known that related halide perovskites exhibit overdamped phonons \cite{fujii1974}, as discussed in various recent experimental and theoretical studies \cite{songvilay2019a,ferreira2020a,stock2020,weadock2020,lanigan-atkins2021b,weadock2023b, fransson2023}.
We perform the same series of static and dynamic calculations as above, now using the PBE functional \cite{perdew1996} with dispersive corrections \cite{tkatchenko2009a} and \texttt{DynaPhoPy} \cite{carreras2017} to compute renormalized phonons and resulting band gaps (see Table~\ref{table:1}).
Importantly, for CPB the band gap computed from MC with renormalized phonons differs by 0.5\,eV from the one obtained in MD.  
Hence, including the effect of renormalized phonons {in a perturbative fashion} when calculating electron-phonon interactions does not capture thermal effects in the electronic structure of CPB well. 
{This motivates our investigation on the impact of overdamped phonons on the fundamental band gap.}
\begin{figure}
    \centering
    \includegraphics[width=0.9\linewidth]{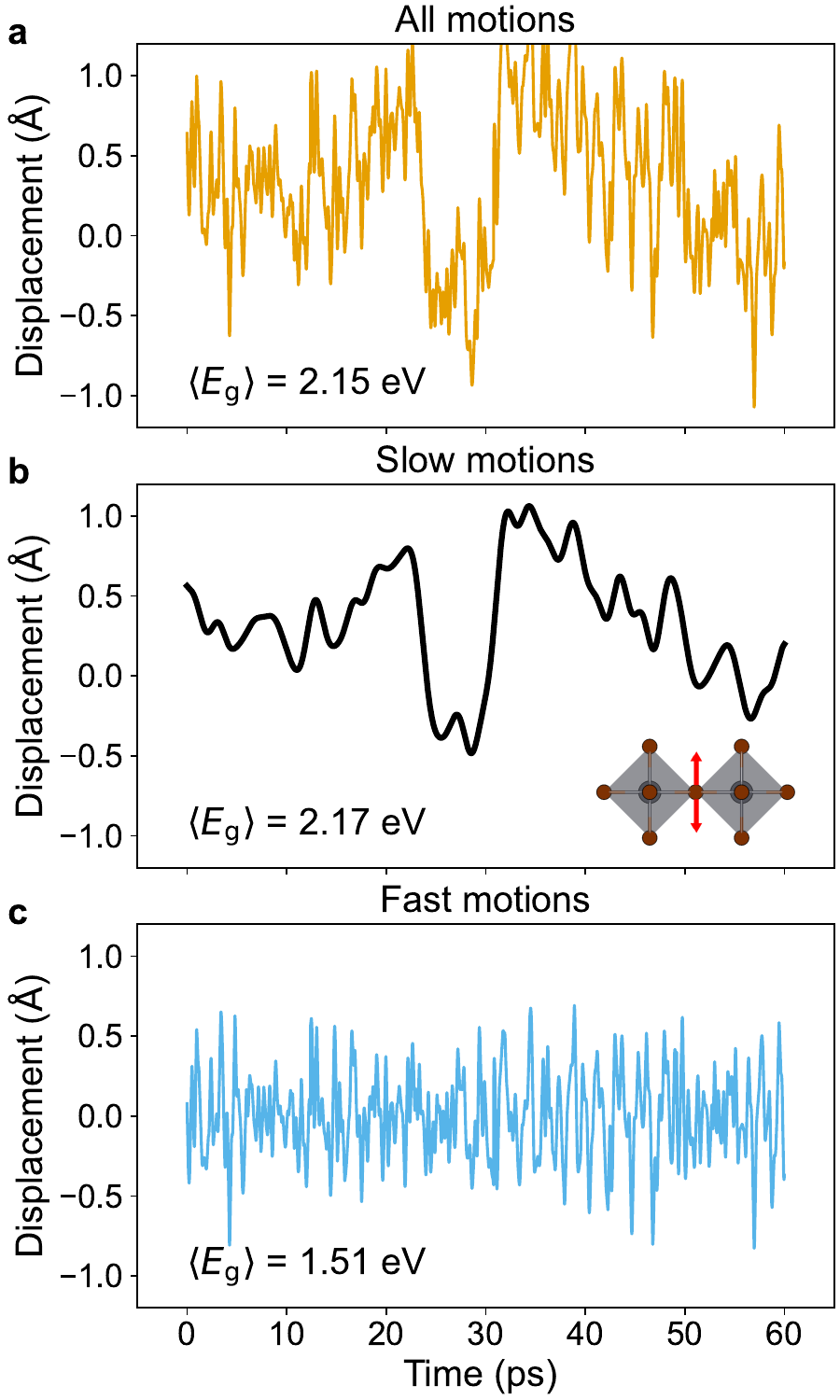}
    \caption{(a) Displacements of Br atoms perpendicular to the Br-Pb-Br axis (see inset in panel b) in MD. (b) Slow components of the trajectory shown in panel a, calculated using Eq.~\ref{eq:2}. (c) Fast components of the trajectory shown in panel a, calculated by subtracting the slow fluctuations (panel b) from full MD (panel a). The insets report the DFT-calculated time-averaged band gap for each trajectory.}
    \label{fig:disp_split}
\end{figure}

{ We investigate the overdamped phonon modes in CPB} using the damped harmonic oscillator (DHO) model which was found to apply to halide perovskites \cite{fransson2023}. 
The overdamped phonons in CPB are soft modes at the M and R point of the Brillouin zone and correspond to octahedral tilting motions \cite{fransson2023}. 
In the DHO model, the velocity autocorrelation function (VACF) of an overdamped system is written as \cite{fransson2021}
\begin{equation}
    C(t) =\frac{A}{\tau_{L}-\tau_{S}}(\frac{1}{\tau_{S}}e^{-t/\tau_{S}}-\frac{1}{\tau_{L}}e^{-t/\tau_{L}}).
    \label{eq:1}
\end{equation}
Here, $A$ is the amplitude of the DHO at $t=0$, and $\tau_L$ and $\tau_S$ are parameters characterizing long- (slow atomic motions) and short-time dynamics (fast atomic motions), respectively.
Hence, given that CPB exhibits overdamped phonons one anticipates two time scales in its atomic dynamics, which were recently found in similar halide perovskite materials \cite{baldwin2024,fransson2023}.

To study the separate impact of the slower and faster atomic dynamics on the band gap, we disentangle the {MD-calculated displacements via a moving average procedure.} 
We calculate time-averaged atomic displacements, $\bar{\mathbf{u}}_i(t)$, as follows:
\begin{equation}
    \bar{\mathbf{u}}_i(t) = \frac{1}{\Delta t}\int_{t-\Delta t}^{t+\Delta t}dt' \mathbf{u}_i(t')w(t'). 
    \label{eq:2}
\end{equation}
Here, $\mathbf{u}_i(t)$ is the MD trajectory of atom $i$, $\Delta t$ is the averaging window, and $w(t)$ is the Blackman window function.
We monitor Br motions perpendicular to Pb-Br-Pb axes (see Fig.~\ref{fig:disp_split}a and inset of panel b) because they were found to be important in various ways \cite{bechtel2019,lanigan-atkins2021b,zhu2022,gehrmann2022b,wiktor2023a}. 
Furthermore, we apply the averaging procedure to Cs displacements as well because the A-site cations also exhibit large displacements in halide perovskites. 

Starting from the full MD trajectory with all motions present (see Fig.~\ref{fig:disp_split}a), and using Eq.~\ref{eq:2} as well as $\Delta t=4$\,ps \cite{fransson2023,baldwin2024}, we obtain the slow motions corresponding to atomic dynamics occurring at longer time scales in CPB (see Eq.~\ref{eq:1} and Fig.~\ref{fig:disp_split}b).
These slower fluctuations are manifestations of dynamical octahedral tiltings between local minima in the multi-well potential energy surface of CPB. 
We calculate the fast motions (Fig. \ref{fig:disp_split}c) by subtracting the trajectory obtained for the slower atomic motions from the full MD trajectory. 
The fast atomic motions correspond to oscillations in the vicinity of local minima in the potential energy surface occurring at shorter time scales. 
We find a band gap for the trajectory of slower atomic motions that is essentially on par with the one from the original MD run, while the band gap corresponding to the trajectory of faster atomic fluctuations is found to be much smaller. 
We can conclude that the large thermal renormalization of the band gap in CPB is due to slow atomic motions in this system.
The effect of slow octahedral tilting motions on the band gap was reported before \cite{quarti2016,whalley2016a,zhao2020,seidl2023}.
{But the presence of two time scales in the atomic dynamics is because of phonon overdamping.
Therefore, our results connect impacts of anharmonic motions on optoelectronic properties of halide perovskites to their phonon band structure, specifically to overdamped phonons.}

\begin{figure}
    \centering
    \includegraphics[width=1\linewidth]{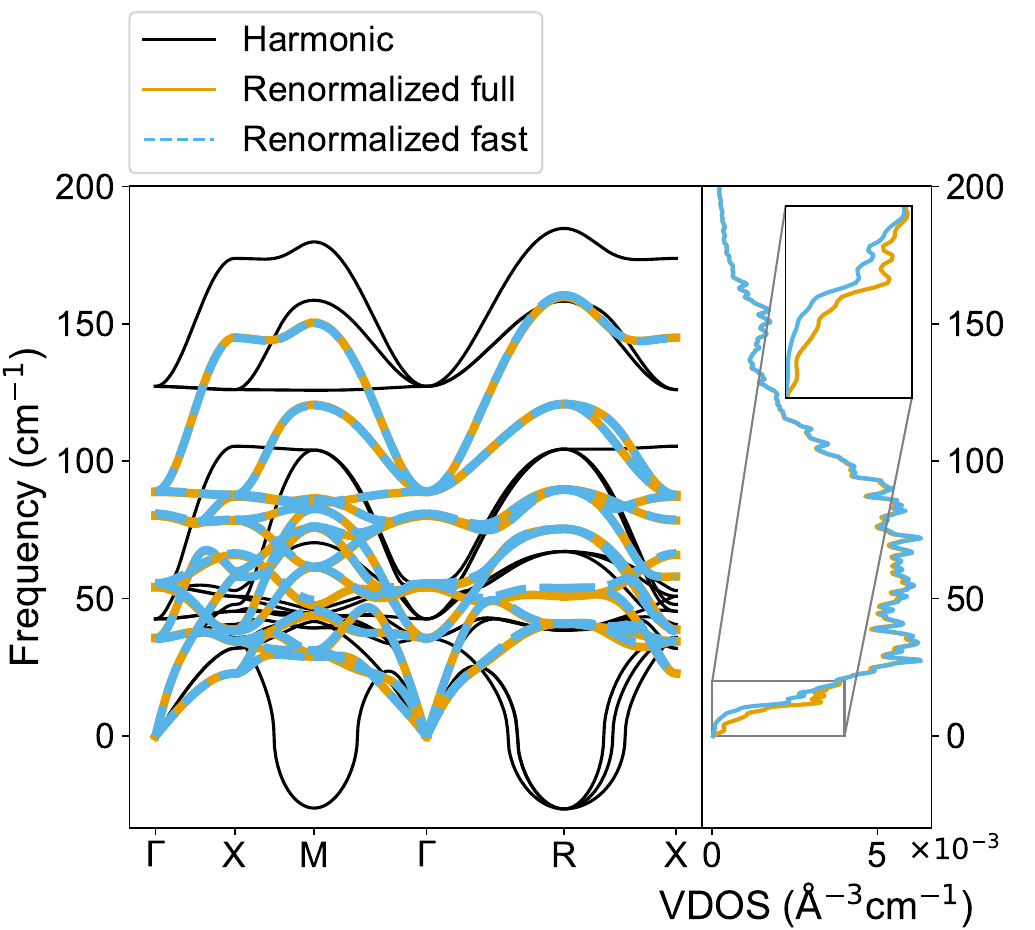}
    \caption{Renormalized phonon dispersion of CPB at 425\,K, calculated using the full MD trajectory (orange line) and the one only including fast fluctuation (blue dashed line), together with the harmonic result (black line). The vibrational density of states (VDOS) of the two differently renormalized systems is shown to the right. The inset provides a zoom into low-frequency regions of the VDOS.}
    \label{fig:ph_split}
\end{figure}

These findings suggest that the breakdown of the phonon quasiparticle picture associated with overdamped phonons has important consequences for the electronic-structure properties at finite temperatures.
We analyze the connection of overdamped phonons and the band gap further by computing renormalized phonon properties with \texttt{DynaPhoPy} for the trajectories obtained for fast atomic motions and full MD, respectively. 
We find that phonon dispersions of CPB at 425\,K are largely independent of whether the fast or full fluctuations are used to renormalize them (Fig. \ref{fig:ph_split}a).
This result might appear surprising but can be rationalized within the DHO model.
For $\tau_{L}\gg\tau_{S}$ in Eq.~\ref{eq:1}, the contribution of slower fluctuations to the VACF is minimal.
Hence, when phonons are overdamped, the VDOS is dictated by the faster atomic dynamics since the vibrational density of states (VDOS) is obtained as a Fourier transform of the VACF. 
Indeed, this is what we find for CPB with the exception of the low frequency range in the VDOS (Fig. \ref{fig:ph_split}b).

{Relevant for the electronic structure, the VACF is also used when computing electron-phonon interactions and band gaps at finite temperatures using renormalized phonons.} 
For example, a normal-mode decomposition analysis of the VACF can be applied to calculate phonon renormalization \cite{carreras2017}. 
In such procedures, the effects of slow atomic dynamics and phonon overdamping on the electronic structure are largely neglected because they are not included in the VACF even in principle.
Hence, when generating structures from the renormalized phonon dispersions with MC for CPB or other overdamped systems, the structures will essentially correspond to the ones of faster atomic motions. 
By contrast, we showed that the slower atomic motions are significantly more relevant than the faster ones for thermal effects in the band gap of CPB (cf. Fig.~\ref{fig:disp_split}).
Therefore, presence of two time scales in a system with overdamped phonons implies that effects of slower atomic motions on the band gap are neglected in standard MC treatments of the electron-phonon interactions. 
{The consequences of overdamped phonons on the electronic structure cannot be captured when including anharmonic effects only implicitly in an effective harmonic model.}

\begin{figure}
    \centering
    \includegraphics[width= 0.8\linewidth]{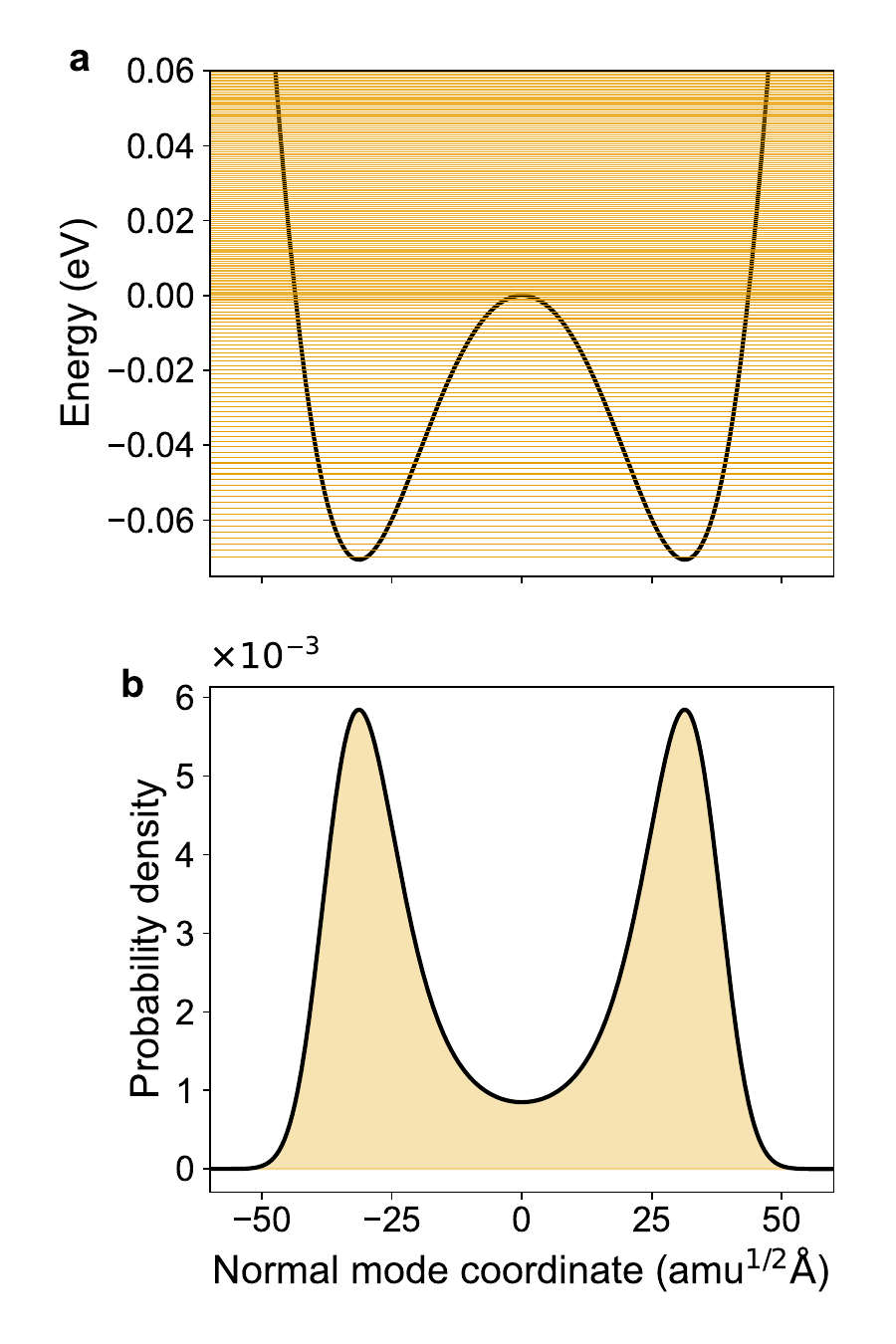}
    \caption{(a) Energy as a function of displacement along the imaginary mode in CPB at the M point of the Brillouin zone together with the eigenvalues of the Schr\"odinger equation for this potential. (b) Probability density distribution for atomic displacements as a function of normal mode coordinate for the potential well of the imaginary mode shown in panel a.}
    \label{fig:SE}
\end{figure}

{We now investigate the effect of accounting for anharmonicity in an MC method by including the imaginary modes of CPB beyond the quadratic potential. 
That is we perform stochastic sampling of the anharmonic region of the potential energy surface in a perturbative method that neglects explicit phonon-phonon interactions.
With it we also investigate the microscopic mechanism of the effect of overdamped phonons on the band gap and connect it to the slow octahedral tilting motions.} 
Therefore, we first solve a Schr\"odinger equation for frozen-phonon potential wells corresponding to the imaginary modes (see Fig.~\ref{fig:SE}a) \cite{skelton2016,whalley2016a,sm2024}. 
This allows us to generate stochastic displacements according to the probability distributions associated with the imaginary modes (see Fig.~\ref{fig:SE}b) and include them in MC.
The octahedral tilting corresponding to the slower atomic dynamics in CPB are then anticipated to be accounted for in a better way compared to the MC procedures discussed above \cite{sm2024}.
We find that this approach yields an improved band gap of 1.92\,eV compared to MC$_\mathrm{harm}$ and MC$_\mathrm{renorm}$, however, it is still roughly 0.2 eV below the one of MD.
Explicit phonon-phonon interactions concurrent with the overdamped phonons in CPB are responsible for this remaining difference and, thus, significantly impact the fundamental band gap.

{We summarize our results and provide guidelines for computing electronic-structure properties at finite temperatures.} 
We found that overdamped phonons can strongly impact the fundamental band gap of perovskites. 
Presence of overdamped phonons leads to two time scales in the atomic motions of which the faster components determine the VACF and properties calculated from it, such as renormalized phonons and resulting band gaps.
However, in agreement with previous work \cite{quarti2016,whalley2016a,zhao2020,seidl2023} we found that the slower atomic motions determine the large thermal renormalization of the band gap in halide perovskites.
{Therefore, we showed that a presence of overdamped phonons leads to changes of the optoelectronic properties that cannot be captured in perturbative treatments of phonon-phonon interactions.}
{This rationale can be understood from our findings of} augmenting MC methods with imaginary modes. {
{Explicitly sampling structures in the anharmonic regions of the potential energy surface led to improved band gaps at 425\,K.
The remaining band-gap difference of 0.2\,eV compared to MD calculations is because the latter non-perturbatively capture phonon-phonon interactions. 
This effect is relevant for the band gap when phonons are overdamped but can be negligible otherwise.}
{Related to what we found here, presence of overdamped phonons required a description going beyond low-order perturbation theory also for thermal transport}\cite{caldarelli2022}.

In conclusion, our work disentangles the consequences of anharmonic effects on the optoelectronic properties and electron-phonon interactions of perovskites. 
The comparison of STO and CPB illustrates the limits of perturbative approaches when predicting band gaps of anharmonic materials.
{The importance of overdamped phonons was discussed in the context of thermoelectric materials recently} \cite{lanigan-atkins2020a,li2023b}.
{However, apart from halide perovskites we are not aware of another semiconductor for optoelectronic applications that exhibits phonon overdamping.}
Here, it is important that we found the slow atomic motions causing large changes of electronic levels to be connected to overdamped phonons, {and likely also to local polar fluctuations observed in Raman} \cite{yaffe2017}.
The large-amplitude oscillations of energy levels occurring on ps time scales have been implicated in the unique optoelectronic properties of halide perovskites, such as their defect tolerance \cite{cohen2019b,wang2022a}. 
Therefore, we believe that search of other semiconductors with similarly overdamped phonons is relevant for future research.

\begin{acknowledgments}
We thank Olle Hellman, Waldemar Kaiser and Sebastián Caicedo-Dávila for fruitful discussions. 
Funding provided by the Deutsche Forschungsgemeinschaft (DFG, German Research Foundation) via Germany's Excellence Strategy - EXC 2089/1-390776260, is gratefully acknowledged. 
The Gauss Centre for Supercomputing e.V. is acknowledged for providing computing time through the John von Neumann Institute for Computing on the GCS Supercomputer JUWELS.
\end{acknowledgments}

%\printbibliography
\bibliographystyle{apsrev4-1}
%%\bibliography{z_revtex_template}% Produces the bibliography via BibTeX.
\bibliography{overdamp2023}
\end{document}